# Angular-Dependent Thermal Hall Effect in a Honeycomb Magnet: Disentangling Kitaev and Dzyaloshinskii-Moriya Interactions

Shuvankar Gupta[#], Olajumoke Oluwatobiloba Emmanuel[#], Pengpeng Zhang, and Xianglin Ke

Department of Physics and Astronomy, Michigan State University, East Lansing, Michigan 48824-2320, USA

#: These authors contributed equally to this work.

Layered honeycomb magnets have garnered significant attention recently for their exotic quantum phenomena due to the potential anisotropic, bond-dependent Kitaev interactions. However, distinguishing the roles of Kitaev interactions and the symmetry-allowed Dzyaloshinskii-Moriya interaction (DMI) remains challenging, since both mechanisms may lead to similar magnetic excitations and thermal transport properties. To tackle this challenge, using a ferromagnetic honeycomb insulator $VI_3$ as a model system, we systematically study the angular-dependent thermal Hall conductivity $K_{xy}(\theta, \Phi)$ with both out-of-plane ($\theta$) and in-plane ($\Phi$) magnetic field rotations. Our results reveal a persistent thermal Hall response for both out-of-plane and in-plane rotating magnetic fields, devoid of the sign-reversal patterns characteristic of Kitaev physics. Instead, quantitative analysis shows that the angular dependent $K_{xy}(\theta, \Phi)$ is governed by the projection between the magnetic moment and a tilted DM vector containing both out-of-plane and in-plane components. These results not only establish the DMI-driven topological magnetic excitations as the origin of the thermal Hall response in $VI_3$ but also highlight the angular-dependent thermal Hall effect measurements as an effective approach for distinguishing competing interactions in quantum magnets.

Probing the thermal Hall effect (THE) is a powerful approach for unraveling the emergent thermal transport properties of quantum materials, where a transverse heat current emerges in the presence of a vertical magnetic field in response to a longitudinal temperature gradient [1,2], an analogy to the electric Hall effect in metallic systems. This phenomenon bridges thermodynamics and quantum mechanics, with the thermal Hall signal carried by quasiparticles such as magnons, phonons, or their hybridization (i.e, magnon-polarons) [1–17]. Importantly, the THE can provide crucial insights into nontrivial topological character and chirality of quasiparticles [1–17]. Magnets with a honeycomb lattice are excellent candidates for investigating the THE owing to their peculiar magnetic excitations resulting from the bond-dependent Kitaev interactions or Dzyaloshinskii-Moriya interaction (DMI) [16–33], both of which are crystalline symmetry-allowed. These interactions may give rise to novel topological magnetic excitations with nontrivial Berry curvature that can lead to the intriguing THE [16–33].

However, disentangling the roles of Kitaev interactions and DMI in honeycomb magnets in determining their magnetic excitations and the resulting THE remains a significant challenge and a hotly debated topic. α-RuCl$_3$ [18,24,28,34] and CrI$_3$ [19,22,23,26,29] are prototype materials of such kind. In the antiferromagnetic α-RuCl$_3$ which has a ground state in proximate to quantum spin liquid (QSL) due to the bond-dependent Kitave interactions, it was originally proposed that the Majorana fermions (fractional spin excitations) lead to a half-integer quantized THE [20,21,30,31,35,36]; alternatively, topological magnons [18,24,28] or phonons [37,38] have been claimed to be the heat carriers accounting for the observed THE features. Similarly, rather than having the anticipated quantized thermal Hall signals of Kitaev QSL, the thermal Hall responses of Na$_2$Co$_2$TeO$_6$, another honeycomb material [39], were reported to be dominated by magnons [40] or phonons [37]. For the honeycomb ferromagnet CrI$_3$, it has been reported that its magnon dispersion measured through inelastic neutron scattering (INS) studies [22], including the observation of magnon gap openings at Dirac (*K*) points, can be nicely simulated using a magnetic Hamiltonian including DMI only [22,23,41,42], or Kitaev interactions [23,43–46], or both [19,47,48]. In addition, it has been shown that the existence of sample mosaics of the co-aligned single crystals used in INS studies can extrinsically introduce an artificial gap feature at *K* points [23,49,50], which further complicates the understanding and disentanglement of DMI and Kitaev interactions in this system. Besides, the recent THE study of $CrI_3$ revealed that the thermal Hall signal anticipated to arise from topological magnons is fairly small if nonzero [26], further casting questions on the nature and the size of Dirac magnon gap.

$VI_3$ [51,52], a sister compound of $CrI_3$, provides a fresh platform to address these challenges. It is a ferromagnetic insulator with $T_C \sim 50$ K proceeded by a structural phase transition around 77 K [51,53]. Single crystal neutron diffraction measurements revealed both in-plane and out-of-plane magnetic moment components, indicating that the magnetic easy-axis is not exactly along the *c*-axis, which is different from $CrI_3$ [54]. Recently, a pronounced anomalous thermal Hall response was observed in $VI_3$, which was argued to be mainly attributed to topological magnons at higher temperature (above 20 K) and to magnon-polarons via magnon-phonon hybridization at lower temperature [25]. The topological magnetic excitations were attributed to the DMI with an out-of-plane DM vector [25]. More recently, an INS study of $VI_3$ claimed that its magnetic excitations are predominately determined by Kitaev interactions [52]. Nevertheless, as described above, it has been shown that similar magnetic excitations of a honeycomb ferromagnet can be nicely constructed using a Hamiltonian with a prominent Kitaev interaction or DMI or both [19,47]. Therefore, the precise role of Kitaev versus DMI interactions in determining the magnetic excitations remains unclear, necessitating more refined techniques to identify the dominant mechanism.

While the previous study [25] largely relied on temperature-dependent THE measurements with an out-of-plane applied magnetic field to infer the nature of topological excitations, the interpretation of the experimental results is often model-dependent and cannot unambiguously distinguish between DMI- and Kitaev-driven mechanisms. In this study, using VI$_3$ as a model system, we address this issue through comprehensive angular-dependent THE measurements by leveraging the magnetic field orientation as a new experimental degree of freedom. We find that the angular dependent thermal Hall conductivity $K_{xy}(\theta, \Phi)$, measured with both out-of-plane ($\theta$) and in-plane azimuthal ($\Phi$) field rotations, retains the same sign under all rotation geometries and is governed by the projection between the magnetic moment and a tilted DM vector containing both out-of-plane and in-plane components. This finding advances the conceptual understanding of the magnetic Hamiltonian with dominating DMI instead of Kitaev interactions that drive the THE in VI$_3$. More broadly, this angle-resolved thermal Hall diagnostic establishes a general framework for probing topological magnetic excitations and resolving intertwined spin–orbit interactions in correlated magnets.

Figure 1(a) presents the temperature dependence of magnetization ($M$) of VI$_3$ measured with magnetic field applied along the $c$-axis. The $M(T)$ data clearly shows a ferromagnetic transition at ~ 50 K, consistent with previous reports [25,51–53]. The bifurcation between the field-cooled (FC) and zero-field-cooled (ZFC) data indicates the formation of magnetic domains when the sample is cooled down in the absence of magnetic field. The top panel in Figure 1(b) shows the isothermal magnetization $M(\mu_0 H)$ data at $T = 37$ K on the same sample (S1) measured at different angles $\theta$ which is defined as the angle between the applied magnetic field relative to the sample in-plane orientation (i.e., $\theta = 90°$ for $H \text{ // } c$-axis and $\theta = 0°$ for $H \text{ // } ab$-plane). One can see that the magnetization along $\theta = 0°$ is notably suppressed compared to $\theta = 45°$ and $90°$ even at 7 T, which suggests that there is an out-of-plane magnetic moment component even in presence of a large magnetic field (7 T) applied along the in-plane direction. This observation indicates a strong magnetic anisotropy. This feature is consistent with recent neutron diffraction studies, which reported that the magnetic moment in VI$_3$ is tilted by approximately 38° degrees from the $c$-axis [54]. In contrast, the bottom panel of Figure 1(b) shows the $M(\mu_0 H)$ curves of the sample S2 measured at $T = 43$ K with the magnetic field applied within the $ab$ plane but along different azimuthal orientations $\Phi$ ($\Phi = 0°$ for $H \text{ // } a$-axis, zigzag direction and $\Phi = 90°$ for $H \text{ // } b$-axis, armchair direction). It is seen that $M(\mu_0 H)$ curves measured at $\Phi = 0°$, $\Phi = 45°$ and $\Phi = 90°$ are nearly identical, suggestive of the absence of in-plane magnetic anisotropy.

Schematics shown in Figures 1(c) and 1(d) illustrate the experimental setups employed for the out-of-plane and in-plane angular dependent thermal transport measurements, respectively, where the definitions of $\theta$ and $\Phi$ are the same as described above. The thermal transport measurements were conducted using a three-sensor-one-heater setup, with the heat current $J_Q$ applied along the zigzag direction and the thermal Hall signal measured along the transverse direction, i.e., armchair direction. Note that since VI$_3$ is moisture-sensitive, all devices were made inside a glove bag with nitrogen gas flowing continuously. Details of the experimental setup are described in the Supplemental Materials and Figure S1 [55]. Figures 1(e) and 1(f) present the temperature-dependent longitudinal thermal conductivity ($K_{xx}$) data for two different magnetic field orientations with $H \text{ // } c$-axis and $H \text{ // } a$-axis, respectively. For both device configurations, the $K_{xx}$ data show similar features with one anomaly observed at ~ 77 K which corresponds to the structural phase transition and another one occurring at ~ 50 K corresponding to magnetic ordering.

At low temperatures (~ 14 K), a broad peak is observed, which results from the interplay between Umklapp phonon scattering and phonon-defect scattering that dominate in different temperature regions. In addition, upon applying a magnetic field of 7 T, $K_{xx}$ is enhanced due to the suppression of phonon scattering by magnons, the population of which is reduced in the presence of magnetic field, consistent with the observation in an early report [25].

Hereafter, we focus on the discussion of angular dependent THE measured around 40 K, at which the THE is mainly driven by topological magnons, to disentangle the roles of Kitaev interactions and DMI in $VI_3$. Figures 2(a)–(d) display the magnetic field dependence of $K_{xy}$ of sample S1 measured at various $\theta$s as the magnetic field is rotated from the in-plane configuration ($\theta = 0°$) to the out-of-plane configuration ($\theta = 90°$) and beyond at $T = 37$ K. The $K_{xy}(\mu_0 H)$ data measured at all $\theta$s are presented in Figure S2 and the isothermal $M(\mu_0 H)$ data measured at various $\theta$s are shown in Figure S3 in the Supplementary Materials [55]. Very intriguingly, one can see that the $K_{xy}(\mu_0 H)$ curves measured at all $\theta$s exhibit a similar field-dependent behavior, even when the magnetic is applied along the in-plane direction as seen in Figure 2(a). That is, $VI_3$ exhibits a sizable planar THE, although the thermal Hall conductivity at $\theta = 0°$ is smaller than that measured at $\theta = 90°$. Figure 2(e) summarizes the variation of $K_{xy}$ as a function of $\theta$ extracted at 1 T at which $K_{xy}$ is nearly saturated. Intriguingly, no sign change in $K_{xy}$ is observed with $\theta$ varying from 0° up to 175°, a significant deviation from the predictions for materials with dominant Kitaev interactions [19,21,30,31,34]. The sign change of THE in systems with dominating Kitaev interactions as the magnetic field rotates between out-of-plane and in-plane orientations (i.e., within the *ac* plane) [19] was proposed to arise from topological phase transition involving a sign change of Chern number of magnons bands or Majorana fermions, with which the thermal Hall conductance is directly associated [20,21,31]. Indeed, sign change of THE is observed in α-RuCl$_3$, a well-studied Kitaev system [20,21,31]. Specifically, in the field-induced QSL, $K_{xy}$ changes sign at critical angles, such as $\theta_1 = \pm 60°$ and $\theta_1 = \pm 45°$ (where $\theta_1$ refers to the angle of the magnetic field relative to the $c$-axis), signifying the angular dependence of the sign of Chern number and the resulting THE [21]. Moreover, it was recently proposed that even in the field-induced polarized state [19,56], the Kitaev-dominating systems also exhibit a similar sign change in $K_{xy}$ as the applied magnetic field rotates within the $ac$ plane. In contrast, no such sign change in $K_{xy}$ is observed in $VI_3$, implying that Kitaev interactions do not play a dominant role in determining the THE in this system. Instead, it points to a fundamentally different mechanism at play in $VI_3$.

To further support the statement above, we also examine the planar THE with the magnetic field direction rotating within the $ab$ plane as illustrated in Figure 1 (d). In the prototype Kitaev material α-$RuCl_3$, it has been proposed that when the heat current ($J_Q$) is parallel to the magnetic field and aligned with the zigzag direction ($a$-axis), the $K_{xy}$ attains a half-quantized value due to the presence of Majorana edge modes in the Kitaev spin-liquid phase [20,21,30,57]. Conversely, when the magnetic field is applied along the armchair direction ($b$-axis), no transverse $K_{xy}$ is observed [20,21]. This behavior is attributed to the preservation of $C_2$ rotational symmetry along the armchair direction, which results in low-energy excitation gap-closing for Majorana fermions. Importantly, with the magnetic field rotating within the $ab$ plane, the Chern number exhibits a six-fold sign (+1 and -1) symmetry. This angular dependence of the Chern number originates from the bond-dependent Kitaev interactions, which couple spin components differently along the three crystallographic directions [21]. As the azimuthal orientation $\Phi$ varies, it induces gap-closing and reopening and modulates the Berry curvature of Majorana fermions bands, leading to sign changes in $K_{xy}$ [20,21,31]. Interestingly, a similar sign structure has been theoretically predicted for the Kitaev-dominating field-polarized states, where magnons are suggested to drive the THE [18,19,58]. Additionally, planar THE has also been reported in another Kitaev candidate compound, $Na_2Co_2TeO_6$ [39], where the planar THE has been attributed to either phonons [37] or topological magnons [40]. Regardless of the mechanism responsible for the planar THEs, the distinctive sign change in $K_{xy}$ by varying $\Phi$ stands for a characteristic feature of Kitaev-dominating systems.

Figures 3(a)-(d) display the magnetic field dependence of planar $K_{xy}$ measured at $T = 43$ K on sample S3 with the magnetic field applied along different in-plane azimuthal ($\Phi$) orientations relative to the $a$-axis (zigzag direction). The complete in-plane azimuthal rotation datasets for samples S2 and S3 are presented in Figures S4 and S5, respectively [55]. As illustrated in Figure 1(d), the thermal gradient $J_Q$ was applied along the zigzag direction and the transverse heat was measured along the armchair direction. Figure 3(e) summarizes the angular dependence of $K_{xy}$ extracted at 1 T for S2 and S3. There are a couple of interesting observations worth pointing out. Firstly, the THE with a sizable magnitude is observed in $VI_3$ even for the configuration with $J_Q$ // $a$-axis and $H$ // $b$-axis (along which the transverse temperature gradient is measured), i.e., at $\Phi = 90°$. This feature is in sharp contrast to the planar THEs reported previously for the Kitaev-dominating α-$RuCl_3$ [20,21] and $Na_2Co_2TeO_6$ [37]. Secondly, the $K_{xy}$ data obtained in $VI_3$ lack the characteristic sign change when varying $\Phi$, as shown in Figure 3(e), a feature drastically different from what is predicted or observed in planar THEs for materials with Kitaev-dominating interactions as discussed above [19–21,31,58]. Therefore, the absence of sign changes in $K_{xy}$ shown in Figure 2(f) and Figure 3(f) when varying $\theta$ and $\Phi$, respectively, conclusively suggests that the topological nature of magnetic excitations in $VI_3$ is not determined by the Kitaev interaction which was recently proposed to be the dominant term in the magnetic Hamiltonian based on INS studies [59].

An intriguing question arises: what mechanism underlies the observed thermal Hall response in $VI_3$? The DMI provides a plausible solution. As discussed previously, similar magnetic excitations of a honeycomb ferromagnet can be equally nicely described using a Hamiltonian with dominant Kitaev interactions or DMI [19,47,58]. The DMI term lifts the degeneracy of magnon bands at the K points, and consequently, such gapped topological magnon bands possess nonzero Berry curvature, as shown in Figure 4(a,b), giving rise to magnon THE [16,17,25]. For ferromagnets with a single type of magnetic ions occupying a perfect honeycomb lattice, the DM vector is oriented along the out-of-the plane (i.e., $c$-axis) due to symmetry constraint; as a result, only the out-of-plane magnetic moment can lead to thermal Hall response. In contrast, for an in-plane magnetic field configuration ($\theta = 0°$, $H // ab$-plane), the thermal Hall signal is anticipated to vanish provided that the magnetic moments are fully aligned within the $ab$-plane and orthogonal to the DMI vector [47]. This is distinct from the thermal Hall response observed in $VI_3$ as shown in Figure 2 and Figure 3, which is presumably associated with the magnetization of $VI_3$. Recent neutron diffraction measurements show that the magnetic moment of $VI_3$ at zero field is tilted approximately 38° away from the $c$-axis [54], and the out-of-plane magnetic moment component preserves even in the presence of 7 T in-plane magnetic field as suggested from the $M(\mu_0 H)$ data (Figure 1(b) and Figure S3 [55]). Consequently, this persistent out-of-plane magnetic moment component, which aligns with the DM vector, can lead to a finite $K_{xy}$, even when $\theta = 0°$. As the magnetic field tilts further towards the out-of-plane configuration (i.e., $\theta$ increases towards 90°), the out-of-plane magnetic moment component increases. This enhanced alignment of out-of-plane magnetic moment with the DMI vector amplifies the thermal Hall signal, resulting in the observed increase in $K_{xy}$ with increasing $\theta$ towards 90° (Figure 2(e)).

Nevertheless, with an out-of-plane DM vector only, one would expect the resulting DMI-driven $K_{xy}$ to be symmetric about $\theta = 90°$ (i.e., $H$ // $c$-axis). This is in contrast to $K_{xy}(\theta)$ presented in Figure 2(e) which shows more complex features: interestingly, $K_{xy}$ maximizes near $\theta = 120°$ instead of 90° and exhibits an asymmetric angular dependence where $K_{xy}$ at $\theta = 175°$ is larger than that at $\theta = 0°$. To understand the observed $K_{xy}(\theta)$ character, one needs to take into account the in-plane DM term. It is known that $VI_3$ undergoes a structural transition from the *R-3* space group to the *P-1* space group below $T_s \sim 77$ K, forming a slightly distorted honeycomb lattice [54,60]. This structural distortion breaks both inversion and mirror symmetries, allowing for both in-plane and out-of-plane DM vector components [44,61–63]. In other words, the DM vector is not simply along the $c$-axis, as illustrated in Figure 4(c). Since the thermal Hall response is directly related to Berry curvature of the gapped magnon bands [14], the latter of which is approximately proportional to the DM strength parallel to the magnetic moment [47,64], it is important to figure out the angular dependence of DMI to understand the measured $K_{xy}(\theta)$. Note that the DMI-driven topological magnon gap is linearly proportional to $\vec{D} \cdot \vec{M}$ [47], that is, to $\cos(\theta_{DM} - \theta_M)$, where $\vec{D}$ stands for the DM vector and $\vec{M}$ represents the magnetic moment vector, and $\theta_{DM}$ and $\theta_M$ define the directions of these two vectors relative to the $a$-axis respectively (Figure 4(c)). With the magnetic field rotated within the $ac$-plane, $\vec{M}$ has components along both $c$-axis and $a$-axis, while for simplicity $\vec{D}$ is assumed to have components along these two directions too. Figure S6 [55] presents the angular dependence of the magnetic moment direction $\theta_M$, which is calculated based on the isothermal $M(\mu_0 H)$ data measured at various $\theta$s shown in Figure S3 [55]. One can see that $\theta_M$ does not simply follow $\theta$ (i.e., the applied magnetic field direction) due to the magnetic anisotropy of $VI_3$ discussed previously. In Figure 2(f), we plot the angular dependence of $\cos(\theta_{DM} - \theta_M)$ with $\theta_{DM}$ assumed to be 100°, which qualitatively agrees nicely with the $K_{xy}(\theta)$ data shown in Figure 2(e). Such a striking agreement suggests that the THE features observed in VI$_3$ are closely correlated with the angle between DM vector and magnetic moment, indicating the DMI-driven instead of Kitaev interactions-driven thermal Hall response in VI$_3$. Figure S7 shows the calculated angular dependence of cos($\theta_{DM}$ - $\theta_M$) with various $\theta_{DM}$ values and the comparison with the normalized $K_{xy}(\theta)$ data. The closeness of $\theta_{DM}$ to 90° implies that the in-plane DM vector is small compared to the out-of-plane DM vector; however, this in-plane DM vector is essential to account for the asymmetric feature (relative to $\theta = 90°$) in $K_{xy}(\theta)$.

The DM vector with a non-zero in-plane component can also qualitatively explain the azimuthal orientation $\Phi$ dependence of $K_{xy}$ presented in Figure 3(e). Since there is no in-plane magnetic anisotropy as shown in Figure 1(b), the out-of-plane magnetic moment component remains invariant when rotating the field within the $ab$-plane, while the in-plane magnetic moment component follows the direction of in-plane magnetic field (Figure 4(d)). As a result, provided that $\theta_{DM} = 90°$, i.e., the DM vector only has an out-of-plane component, one would expect $K_{xy}$ to be independent of $\Phi$. Nevertheless, we do see small variation in $K_{xy}$ as $\Phi$ changes, as shown in Figure 3(e). To quantify this, we calculate the $\Phi$ dependence of normalized $\vec{D} \cdot \vec{M}$ (i.e, $\vec{D} \cdot \vec{M}/(|D| * |M|)$), which is equal to $\sin(\theta_{DM})\sin(\theta_M) + \cos(180 - \theta_{DM})\cos(\theta_M)\cos(\Phi)$. The result is shown in Figure 3(f), which shows slight dependence of $\Phi$, qualitatively consistent with the $K_{xy}(\Phi)$ feature shown in Figure 3(e). This again affirms that the $K_{xy}$ is directly related to the DM strength parallel to the magnetic moment, further demonstrating that the THE in VI$_3$ is associated with the topological magnons mainly driven by the DMI instead of Kitaev interactions. Moreover, our finding of the DM vector with both in-plane and out-of-plane components calls for future studies, such as $ab$ initio calculations and/or revisit of the inelastic neutron scattering studies, to further examine the DM vectors in this honeycomb magnet.

In summary, via comprehensive angular dependent thermal Hall effect measurements on VI$_3$, we show that the obtained thermal Hall conductivity $K_{xy}(\theta, \Phi)$ does not exhibit sign reversals anticipated for Kitaev physics when varying either out-of-plane ($\theta$) or in-plane azimuthal ($\Phi$) rotation. Instead, our quantitative analysis of the angular dependent $K_{xy}$ can be elegantly captured by the projection between the magnetic moment and the tilted DM vector with both out-of-plane and in-plane components. These findings unambiguously demonstrate that the THE in VI$_3$ is induced by the topological magnons driven by the DMI instead of Kitaev interactions, which advances the conceptual understanding of the magnetic Hamiltonian in this system. More broadly, this work represents an excellent example of using angle-resolved THE to disentangle Kitaev interaction and DMI contributions to magnetic excitations of quantum materials with competing terms in the Hamiltonian.

Data for the figures presented in this paper are available at https://doi.org10.5281/zenodo.17991096 [65].

S. G. and X.K. acknowledge financial support from National Science Foundation (DMR-2219046). O.E., and the SQUID measurements were supported by the U.S. Department of Energy, Office of Science, Office of Basic Energy Sciences, Materials Sciences and Engineering Division under Grant No. DE-SC0019259. P. P. Zhang acknowledges the financial support from the U.S. Department of Energy, Office of Basic Energy Sciences, Division of Materials Sciences and Engineering under Award Number DE-SC0019120.

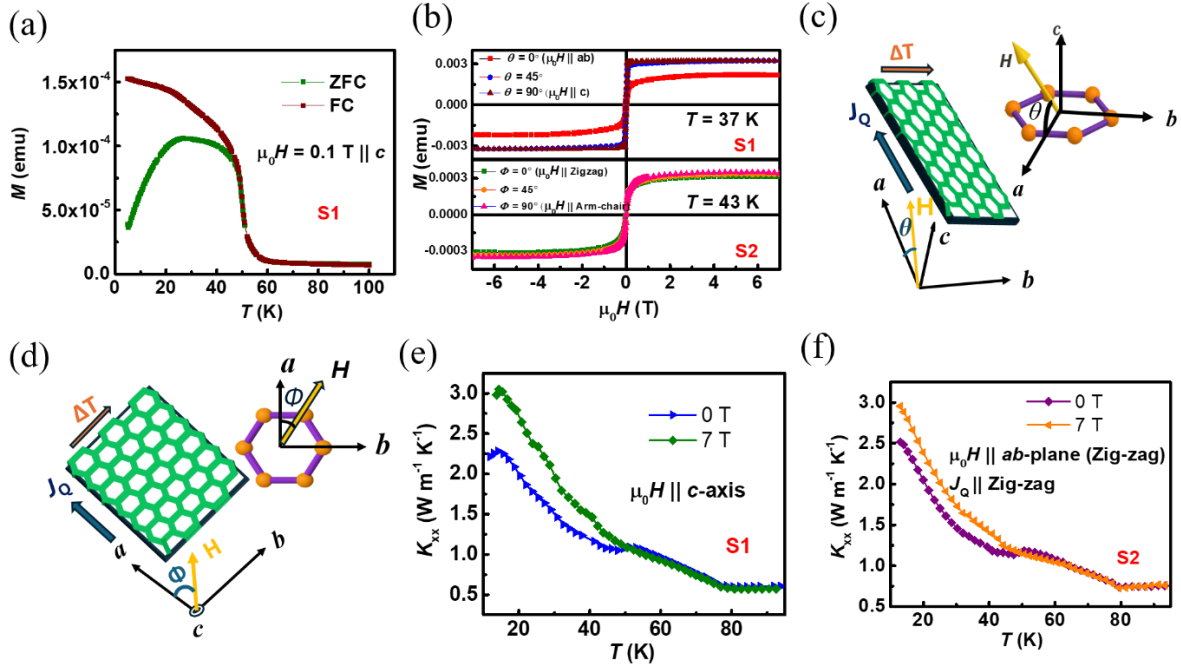

**Figure 1**: (a) Temperature-dependent magnetization ($M(T)$) measured at 0.1 T with $\mu_0 H$ // $c$-axis for sample S1, showing ferromagnetic ordering near $T_C \sim 50$ K. (b) Isothermal magnetization ($M(\mu_0 H)$) for out-of-plane to in-plane ($\theta$-rotation, upper) and in-plane (azimuthal rotation ($\Phi$), lower) field configurations for samples S1 and S2, respectively. (c, d) Schematic experimental setup for thermal Hall measurements: $J_Q$ // $a$-axis (zigzag direction) with transverse heat detected along $b$-axis (armchair direction). In the $ac$-plane (c), $\theta = 0°$ and 90° correspond to $\mu_0 H$ // $a$-axis and $c$-axis, respectively. In the $ab$-plane (d), $\Phi = 0°$ and 90° correspond to $\mu_0 H$ // $a$-axis (zigzag direction) and $b$-axis (armchair direction), respectively. (e, f) Temperature dependence of longitudinal thermal conductivity ($K_{xx}$) for $\mu_0 H = 0$ T and 7 T with (e) $\mu_0 H$ // $c$-axis (sample S1) and (f) $\mu_0 H$ // $a$-axis (zigzag direction) (sample S2).

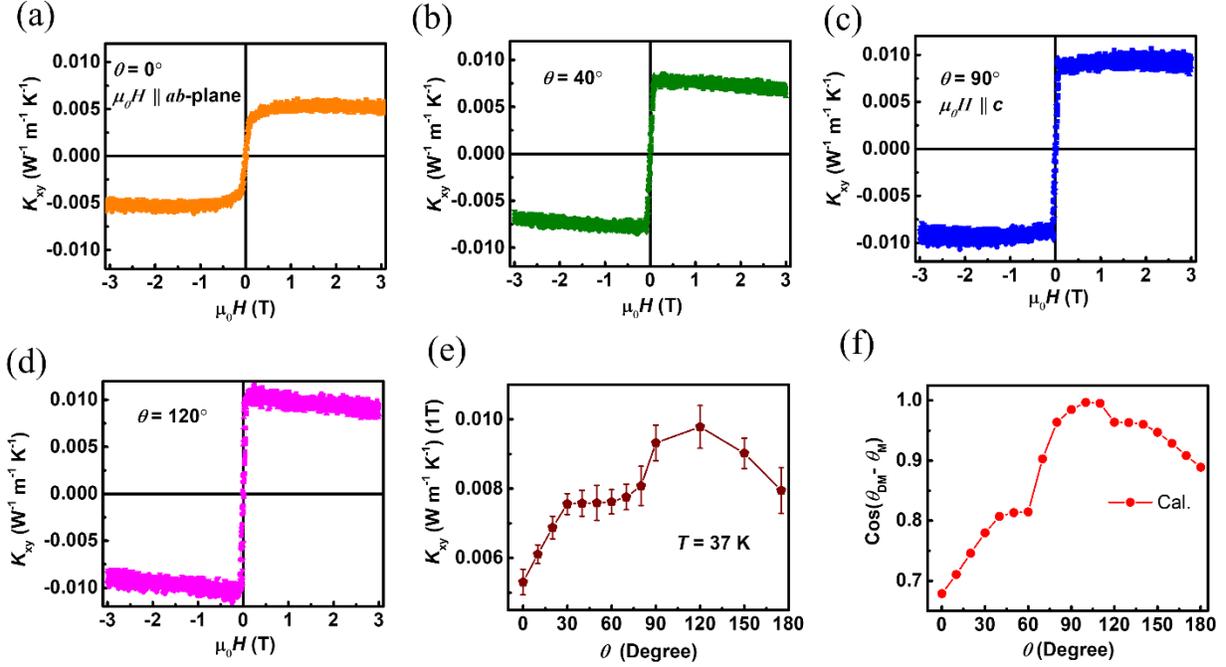

**Figure 2:** Magnetic field dependence of the thermal Hall conductivity ($K_{xy}$) measured at 37 K for sample S1 at various magnetic field rotations relative to the *ab*-plane: (a) $\theta = 0°$ ($\mu_0 H$ // a-axis) (b) $\theta = 40°$ (c) $\theta = 90°$ ($\mu_0 H$ // c-axis) and (d) $\theta = 120°$. (e) $\theta$-dependence of $K_{xy}$ measured at 37 K, highlighting the angular dependence (in-plane to out-of-plane) of the thermal Hall response (f) $\theta$-dependence of $\cos(\theta_{DM} - \theta_M)$, where $\theta_{DM}$ and $\theta_M$ define the directions of the **DM** and **M** vectors relative to the *a*-axis, respectively.

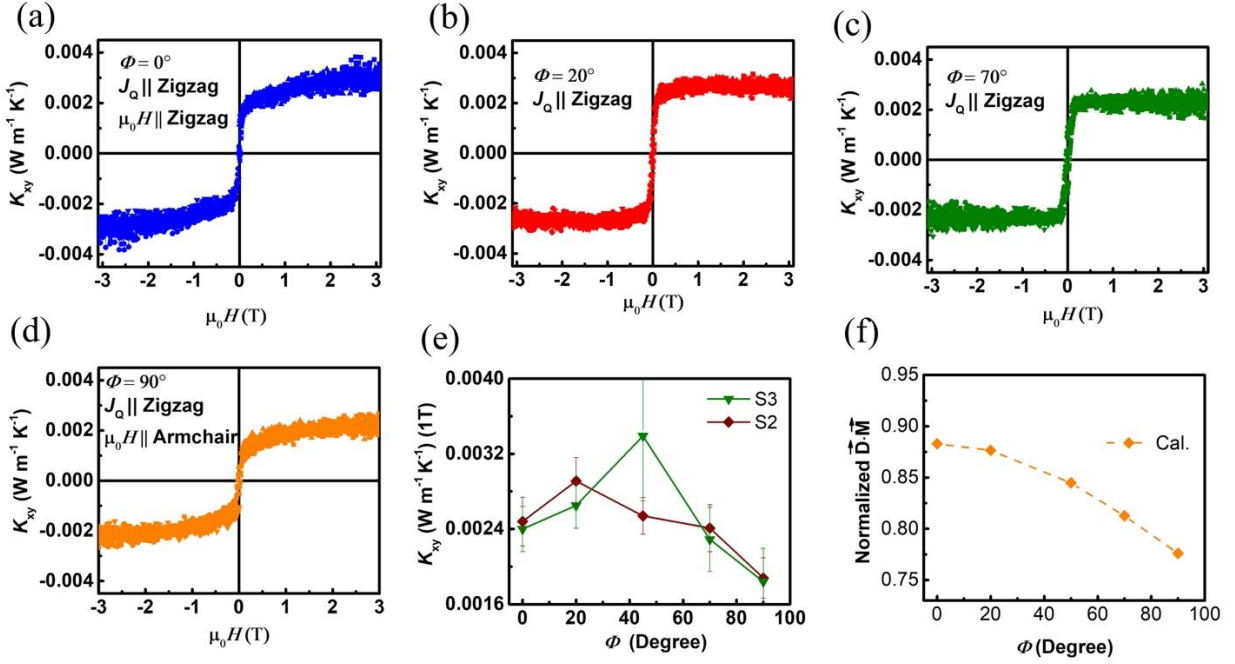

**Figure 3:** Magnetic field dependence of the planar thermal Hall conductivity ($K_{xy}$) measured at 43 K for sample S3 at various azimuthal rotations: (a) $\Phi = 0°$ ($\mu_0H$ // a-axis (zigzag direction)) (b) $\Phi = 20°$ (c) $\Phi = 70°$ and (d) $\Phi = 90°$ ($\mu_0H$ // b-axis (armchiar direction)). (e) $\Phi$-dependence of $K_{xy}$ measured at 43 K, highlighting the angular dependence of the planar thermal Hall response. (f) $\Phi$ dependence of normalized $\vec{D} \cdot \vec{M}$ (i.e, $\vec{D} \cdot \vec{M}/(|D| * |M|)$) as described in the text.

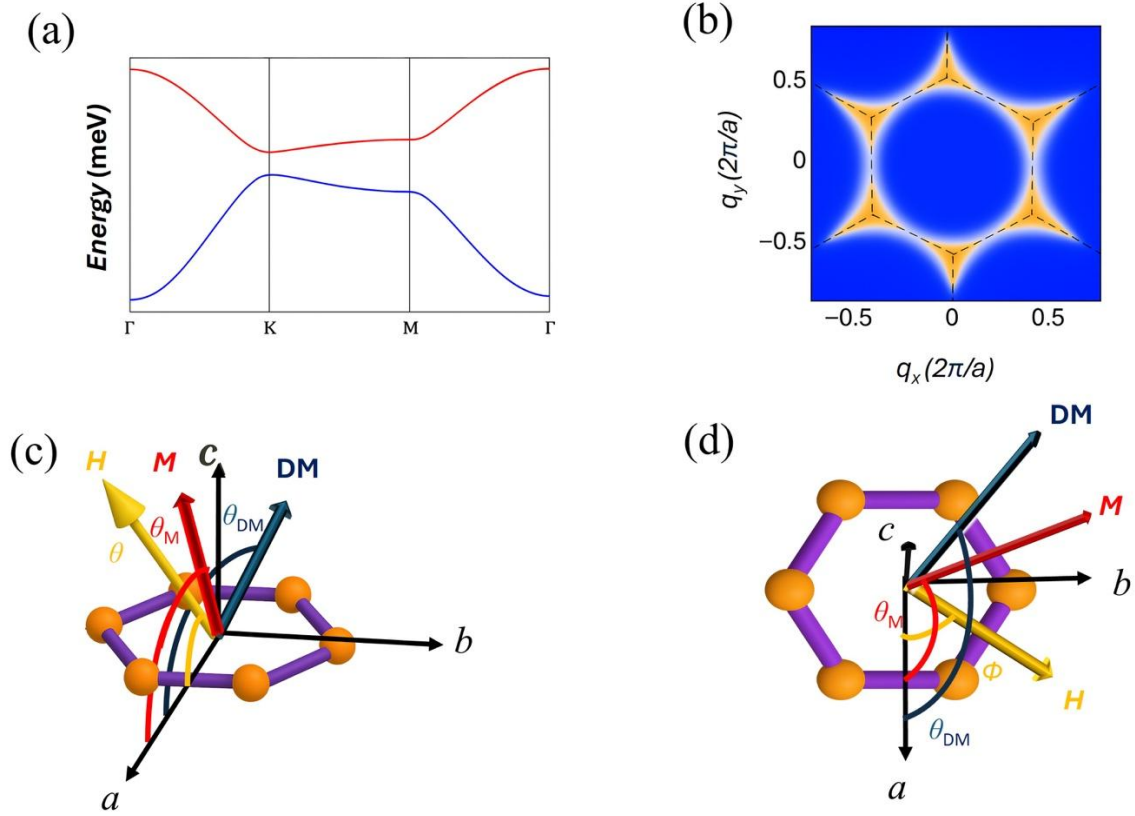

**Figure 4:** (a) Schematic magnon bands showing topological gap opening at K-points. (b) Berry curvature associated with topological magnon bands. (c) Schematic illustration of magnetic moment (*M*), external field (*H*), and **DM** vector for the out-of-plane to in-plane measurement configuration ($\theta$-variation). The *M*, *H* and DM are oriented at respective angles $\theta_M$, $\theta$, and $\theta_{DM}$ relative to the crystallographic *a*-axis. (d) Schematic illustration of *M*, *H* and **DM** vectors for in-plane measurement configuration where $\Phi$ is the angle between *H* and *a*-axis.